\UseRawInputEncoding %for tex live

\documentclass[aps,pra,twocolumn,showpacs,superscriptaddress]{revtex4-1}
\usepackage{graphicx}% Include figure files
\usepackage{dcolumn}% Align table columns on decimal point
\usepackage{bm}% bold math
\usepackage{color}
\usepackage{times}
\usepackage{amssymb}
\usepackage{amsmath}
\usepackage{epsfig}
\usepackage{epstopdf}
\usepackage{dsfont}
\usepackage{subfigure}
\usepackage[colorlinks=true,citecolor=blue, linkcolor=blue,urlcolor=blue]{hyperref}
\usepackage{setspace}
\usepackage{bookmark}
\usepackage{color}

\newenvironment{proof}{{\indent  \it Proof:\,}}{\hfill $\blacksquare$\par}
\definecolor{Green3}{rgb}{0.80,0.87,0.76}
\begin{document}
%\preprint{}

\title{Unification of quantum resources in tripartite systems}

\author{Dong-Dong Dong}
 \affiliation{School of Physics and Optoelectronics Engineering, Anhui University, Hefei
230601,  People's Republic of China}
 \author{Geng-Biao Wei}
  \affiliation{School of Physics and Optoelectronics Engineering, Anhui University, Hefei
230601,  People's Republic of China}
\author{Xue-Ke Song}%
 \email{songxk@ahu.edu.cn}
 \affiliation{School of Physics and Optoelectronics Engineering, Anhui University, Hefei
230601,  People's Republic of China}
 \author{Dong Wang}
 \email{dwang@ahu.edu.cn}
  \affiliation{School of Physics and Optoelectronics Engineering, Anhui University, Hefei
230601,  People's Republic of China}
 \author{Liu Ye}
  \affiliation{School of Physics and Optoelectronics Engineering, Anhui University, Hefei
230601,  People's Republic of China}


\date{\today}

\begin{abstract}
In quantum resource theories (QRTs), there exists evidences of intrinsic connections among different measures of quantum resources, including entanglement, coherence, quantum steering, and so on. However,
building the relations among different quantum resources is a vital yet challenging task in multipartite quantum systems. Here, we focus on a unified framework of interpreting
the interconversions among different quantum resources in tripartite systems.
In particular, an exact relation between the generalized geometric measure and the
genuinely multipartite concurrence are derived for tripartite entanglement states.
Then we obtain the tradeoff relation between the first-order coherence and the genuine tripartite entanglement by the genuinely multipartite concurrence and concurrence fill.
Furthermore, the tradeoff relation between the maximum steering
inequality violation and concurrence fill for an arbitrary three-qubit pure state is found. In addition, we investigate the close relation between the
maximum steering inequality violation and the first-order coherence. The results show that these quantum resources are intrinsic related and can be converted to each other
in the framework of QRTs, although they are still regarded to be different.
\end{abstract}

%\keywords{}

\maketitle

\section{INTRODUCTION}
Entanglement, coherence and steering are known as three vital physical resources in quantum-information processing from the perspective of QRTs \cite{PhysRev.47.777,nielsen_chuang_2010,schrodinger_1935,RevModPhys.91.025001}. Quantum entanglement is a distinctive and unique feature of quantum mechanics, which is obviously different from those of classical mechanics. The quantification of entanglement is a crucial topic in quantum information science. A number of
measures of multipartite entanglement have been put forth,
such as the concurrence \cite{PhysRevLett.80.2245}, entanglement of formation \cite{doi:10.1080/09500349908231260}, generalized geometric measure (GGM) \cite{PhysRevA.81.012308,PhysRevA.94.022336,PhysRevA.95.022301,PhysRevA.68.042307}, genuinely multipartite concurrence (GMC) \cite{PhysRevA.83.062325,PhysRevA.86.062303}, and concurrence fill \cite{PhysRevLett.127.040403}. On the other hand, coherence is an essential property of quantum physics \cite{10.1143/PTP.69.80}, which can show the traits of a stream of photons \cite{mandel_wolf_1995}.
It plays a key role in various quantum algorithms and quantum communication protocols \cite{RevModPhys.94.015004,giovannetti2011advances}. Also it is the main reason why quantum tasks can be realized faster than classical ones \cite{Ma_2019}.
It is based on the superposition principle, which is also the essence of entanglement. Since both
of them can be quantified and characterized by the QRTs, it is reasonable to investigate whether they can be quantitatively converted \cite{PhysRevLett.94.173602,https://doi.org/10.1002/qute.202100036,PhysRevA.89.052302,PhysRevLett.115.220501,PhysRevA.97.042305,PhysRevA.95.062324,Fan_2019}.

%While these measures of quantum entanglement are still regarded differently so far, there exists many indications showing that they are intrinsic related and

The concept of steering was first
introduced by Schr{\"o}dinger for the bipartite case \cite{schrodinger_1935}.
Steering presents a peculiar phenomenon of quantum physics that the local manipulation of one particle of the entangled state can steer another one in the distance, which is not feasible if the two particles are only classically correlated. The quantum steering can be verified by the violation of various steering inequalities, including the linear steering criterion \cite{PhysRevA.80.032112,PhysRevA.93.020103}, the steering
criterion from geometric Bell-like inequality \cite{PhysRevA.91.032107}, the steering criteria from entropic uncertainty relations \cite{PhysRevLett.106.130402,PhysRevA.87.062103,PhysRevA.98.062111,https://doi.org/10.1002/andp.201900124}, and so on. %\cite{PhysRevA.40.913,,chen2013all,PhysRevLett.115.210401,Cavalcanti:15,Jevtic:15}
The steerable states are a strict subset of the entangled
states and a superset of Bell nonlocal states \cite{PhysRevA.92.032107}.
 Steerable states are shown to have many potential applications in randomness generation \cite{Law_2014}, subchannel
discrimination \cite{PhysRevLett.114.060404}, quantum information processing \cite{PhysRevLett.107.020401}, and one-sided device-independent processing in quantum key
distribution \cite{PhysRevA.85.010301}. Recently, the investigations related to the quantum steering have
attracted considerable attention in both theory and experiment \cite{PhysRevA.99.030101,PhysRevLett.125.020404,PhysRevA.103.022207}. For instance, in 2016, Costa \emph{et al.} derived
closed formulas to quantify the linear steering of two-qubit states \cite{PhysRevA.93.020103}.
In 2019,
Pramanik \emph{et al.} experimentally revealed the hidden quantum steerability by using local filtering operations \cite{PhysRevA.99.030101}.
In 2020, Wollmann \emph{et al.} analyzed and experimentally demonstrated quantum steering using generalized entropic criteria and
dimension-bounded steering inequalities \cite{PhysRevLett.125.020404}.

%In the last couple of years, the correlation
%statistics of bipartite subsystems has been proved to be quite effective in studying multipartite composite systems \cite{PhysRevLett.108.110501,PhysRevA.86.042339,doi:10.1126/science.1247715,PhysRevA.71.010301,PhysRevLett.94.153601,PhysRevLett.99.250405,PhysRevA.87.034301}.

While these measures of quantum resources are still regarded differently so far, there exists many indications showing that they are fundamentally connected \cite{PhysRevLett.115.020403,PhysRevA.92.022112,PhysRevA.97.042110,PhysRevA.97.012331,PhysRevA.102.052209,Ding:21,PhysRevA.103.032407,PhysRevA.105.022425}. In 2015, Streltsov \emph{et al.} showed that any degree of coherence in regard to some reference basis can be converted to entanglement \cite{PhysRevLett.115.020403}.
The shareability of three-setting linear steering and its relations with bipartite or tripartite entanglement of three-qubit states were investigated by Paul \emph{et al.} \cite{PhysRevA.102.052209}.
In 2021, Ding \emph{et al.} given the experimental verification of the relationship between first-order coherence and linear steerability  in all-optical systems \cite{Ding:21}.
More recently, Dai \emph{et al.} presented a further study on the complementary relations between tripartite entanglement
and the reduced bipartite steering for three-qubit states \cite{PhysRevA.105.022425}.
However, it is worth noting that most of the related
studies are concerning the two-qubit systems or reduced bipartite subsystems of tripartite systems. Little
attention are paid to the whole entangled multipartite systems. In fact, the investigations on the relation among genuine measures of quantum resources in tripartite systems are important to
understand how the information transfer and flow in the framework of QRTs.

In this paper, we establish a unification of three tripartite measure of quantum resources, including quantum entanglement, coherence, and quantum steering, in tripartite entanglement states.
Firstly, we establish an exact functional relation between the GGM and the GMC for three-qubit pure states.
Then, the tradeoff relation between the first-order coherence and the genuine tripartite entanglement quantified by the GMC and concurrence fill are found.
In addition, we find that there exists a tradeoff relation between the maximum steering
inequality violation and concurrence fill for tripartite pure states. Moreover, we present the close relation between the
maximum steering inequality violation and the first-order coherence. Note that the boundary states of all the above relations consist of the three states: $|\psi \rangle _\alpha$, $|\psi \rangle _m$, and $|\psi \rangle _\theta$. These relations among different measures of quantum resources provide the evidence that different quantum resources are interconnected and can be converted to each other.

This paper is organized as follows: In Sec. \ref{sec2}, we briefly
review some measures of quantum resources in QRTs.
In Sec. \ref{sec3}, we present the tradeoff relations between the genuine tripartite entanglement and the first-order coherence. In Sec. \ref{sec4}, we
study the tradeoff relation between the maximum steering inequality violation and concurrence fill.
The close relation between the maximum steering inequality violation and first-order coherence is derived in Sec. \ref{sec5}.
A summary is provided in Sec. \ref{sec6}.

\section{PRELIMINARIES\label{sec2}}
Here, we give a brief view of the different measures of nonclassicality in QRTs, including
entanglement, coherence, and steering inequality. For multipartite systems, we use GGM, GMC, and concurrence fill as the measures of genuine tripartite entanglement,
which have already been generated and verified by experiments \cite{doi:10.1126/sciadv.aar3931,PhysRevX.8.021012,saggio2019experimental}. The coherence and steering inequality are quantified by first-order coherence and the three-setting linear steering inequality, respectively.

%If a multipartite quantum state cannot be separable in any bipartite split, we will call it a genuinely entangled state.
%The measures of genuine tripartite entanglement employed later are the GGM, GMC, and concurrence fill.
%While the first two measures are defined by the concept of the minimal
%entanglement among all feasible bipartitions, the third one is based on the idea of
%triangle measure.
%The first-order coherence we choose is extensively used in
%the optical systems and easier to experiment with.

\subsection{GGM}
The GGM, as a generalization of the
measure defined by Wei and Goldbart \cite{PhysRevA.68.042307},
is based on the geometric distance between the $n$-partite
state $\left | \psi  \right \rangle$ and the set of all multiparty states $\left | \varphi  \right \rangle$ that are not
genuinely entangled. That is,
\begin{align}
\mathcal{G}(|\psi \rangle ) = 1 - \mathop {\max }\limits_{|\varphi \rangle } |\langle \varphi \mid \psi \rangle {|^2},
\label{Eq.GGM1}
\end{align}
where the maximization is done over all separable states $\left | \varphi  \right \rangle$. An equivalent mathematical expression
of the GGM is given by
\begin{align}
{\cal G}(|\psi \rangle )\! = \!1\! - \!\max \left\{ {\lambda _{I:L}^2\!\!\mid\!\! I \cup L\! =\! \{ 1,2, \ldots ,n\} ,\!I \cap L\! =\! \emptyset } \right\},
\label{Eq.GGM2}
\end{align}
where ${\lambda _{I:L}}$ is the maximal Schmidt coefficient in the I : L
split of the state $\left | \psi  \right \rangle$. For the arbitrary pure states, the $\lambda _{I:L}^2$ are equal to the corresponding eigenvalues of the reduced density matrices ${\rho _I}$ as well as $\rho _L$.

\subsection{GMC}
For multipartite pure states, Ma \emph{et al.} \cite{PhysRevA.83.062325} defined the GMC  satisfying the necessary conditions for being a
multipartite entanglement measure. It is related to the entanglement of the minimum bipartite linear entropies, instead of von Neumann entropies.
For an $n$-partite pure state $\left | \psi  \right \rangle
  \in {{\cal H}_1} \otimes {{\cal H}_2} \otimes  \cdots  \otimes {{\cal H}_n}$
  with dim$\left(\mathcal{H}_{i}\right)=d_{i},i=1,2, \ldots, n$, the GMC is defined as
  \begin{align}
\mathcal{C}(|\psi\rangle)=\min _{\mu_{i}} \sqrt{2\left[1-\operatorname{Tr}\left(\rho_{A_{\mu_{i}}}^{2}\right)\right]},
\label{Eq.GMC1}
\end{align}
where ${\mu _i}$ donates the elements in the set of all feasible
bipartitions $\left\{ {{A_i}|{B_i}} \right\}$. The GMC can be generalized to mixed states $\rho$ via the convex roof construction
  \begin{align}
\mathcal{C}(\rho)=\inf _{\left\{p_{i},\left|\psi_{i}\right\rangle\right\}}
\sum_{i} p_{i} \mathcal{C}\left(\left|\psi_{i}\right\rangle\right),
\label{Eq.GMC2}
\end{align}
where the infimum is over all feasible decompositions
$\rho  = \sum\limits_i {{p_i}} \left| {{\psi _i}} \right\rangle \langle {\psi _i}|$.

\subsection{Concurrence fill}
For tripartite entanglement states, concurrence fill is introduced as a faithfully genuine entanglement measure, based on the area of an alleged
concurrence triangle \cite{PhysRevLett.127.040403}. In the proposal, the lengths of the three sides are set equal to the squares of the three bipartite concurrences.
From Heron's formula for triangle area, the concurrence fill can be defined as
 \begin{align}
 {\cal F}(|\psi \rangle )\!\!=\!\!\left[\!\frac{16}{3} \!Q\!\left(\!Q\!-\!C_{A(\!BC\!)}^{2}
 \!\right)\!\!\left(\!Q\!-\!C_{B(\!AC\!)}^{2}\!\right)\!\!\left(\!Q\!-\!C_{C(\!AB\!)}^{2}\!\right)\!\right]^{1 / 4},
\label{Eq.F123}
\end{align}
where
 \begin{align}
Q=\frac{1}{2}\left(C_{A(BC)}^{2}+C_{B(AC)}^{2}+C_{C(AB)}^{2}\right).
\label{Eq.Q}
\end{align}
$Q$ is the half-perimeter, which is equivalent to the global
entanglement \cite{doi:10.1063/1.1497700,brennen2003observable}. The coefficient $16/3$ guarantees the normalizing condition that $0 \le {{\cal F}_{123}} \le 1$, and the extra square root exceeding
Heron¡¯s formula ensures local monotonicity under the local
quantum operations assisted with classical communications. The $C_{i(jk)}$ can be calculated as following \cite{PhysRevA.61.052306}:
\begin{align}
C_{i(jk)}^{} = 2\sqrt {\det {\rho _i}} ,
 \label{Eq.C}
\end{align}
where $i,j,k \in \{ A,B,C\} $, $i \ne j \ne k$, $\rho _i$ is the reduced density
matrices of the quantum state $\rho _{ABC}$.
It can be found that $0 \le {C_{i(jk)}} \le 1$.
Concurrence fill can detect the difference between entanglements of some states,
while other genuine multipartite entanglement measures can not. In particular, for three-qubit systems, the GMC is equal to the square root of the shortest side length of the concurrence triangle.

\subsection{First-order coherence}
For the three-qubit state $\rho_{ABC}$, the first-order coherence for each subsystem A, B or C is defined by its purity \cite{mandel_wolf_1995}
\begin{align}
{\cal D}\left(\rho_{i}\right)=\sqrt{2 \operatorname{Tr}\left(\rho_{i}^{2}\right)-1},
\label{Eq.D1}
\end{align}
where $i \in \{ A,B,C\} $.
When all subsystems are regarded as independently, the first-order coherence for the state $\rho_{ABC}$ is given by \cite{PhysRevLett.115.220501}
\begin{align}
{\cal D}\left(\rho_{A B C}\right)=\sqrt{\frac{{\cal D}\left(\rho_{A}\right)^{2}+{\cal D}\left(\rho_{B}\right)^{2}+{\cal D}\left(\rho_{C}\right)^{2}}{3}},
\label{Eq.Dabc}
\end{align}
where $0 \le {\cal D}({\rho _{ABC}}) \le 1$. Note that, the first-order coherence is independent of the selection of the reference
basis.

\subsection{The three-setting linear steering inequality violation}
 Cavalcanti \emph{et al.} \cite{PhysRevA.80.032112} formulated the following linear steering inequalities
to verify whether a bipartite state is steerable from Alice to
Bob when both of them are enable to operate $n$ dichotomic
measurements on their own subsystems:
\begin{align}
F_{n}\left(\rho_{A B}, \mu\right)=\frac{1}{\sqrt{n}}\left|\sum_{k=1}^{n}\left\langle A_{k} \otimes B_{k}\right\rangle\right| \leqslant 1,
\label{Eq.Fn}
\end{align}
where $A_{k}=\hat{a}_{k} \cdot \vec{\sigma}$ and $B_{k}=\hat{b}_{k} \cdot \vec{\sigma}$
with $\vec{\sigma}=\left(\sigma_{1}, \sigma_{2}, \sigma_{3}\right)$ being
the Pauli matrices, $\hat{a}_{k}, \hat{b}_{k} \in \mathbb{R}^{3}$
are unit and orthonormal vectors, $\,\left\langle A_{k} \otimes B_{k}\right\rangle=\operatorname{Tr}\left(\rho_{A B}\left(A_{k} \otimes B_{k}\right)\right)$,
and $\mu=\left\{\hat{a}_{1}, \hat{a}_{2}, \ldots, \hat{a}_{n}, \hat{b}_{1}, \hat{b}_{2}, \ldots, \hat{b}_{n}\right\}$
is the set of measurement directions. \\
\indent In the Hilbert-Schmidt representation, any two-qubit state can be
expressed as
\begin{align}
\rho_{A B}\!=\!\frac{1}{4}\!\left[I_{2} \otimes I_{2}\!+\!\vec{a} \cdot \vec{\sigma} \otimes I_{2}\!
+\!I_{2} \otimes \vec{b} \cdot \vec{\sigma}\!+\!\sum_{i, j} t_{i j} \sigma_{i} \otimes \sigma_{j}\right],
\label{Eq.rhoAB}
\end{align}
where $\vec{a}$ and $\vec{b}$ are the local bloch vectors,
 ${t_{ij}} = {\mathop{\rm Tr}\nolimits} \left( {{\rho _{AB}}\left( {{\sigma _i} \otimes {\sigma _j}} \right)} \right)$,
 and ${T_{AB}} = [{t_{ij}}]$ is the
correlation matrix. For the three measurement settings, the state $\rho _{AB}$ is $F_3$ steerable if and only if \cite{PhysRevA.102.052209}
\begin{align}
{\cal S}_{A B}=\operatorname{Tr}\left(T_{A B}^{T} T_{A B}\right)>1,
\label{Eq.Sab}
\end{align}
where the superscript $T$ represents the transpose of the correlation matrix $T_{AB}$.
 Among the three bipartite reduced
states of a three-qubit state ${\rho _{ABC}}$, ${{\cal S}_{\max }}({\rho _{ABC}})$ is defined as the one with the maximum steering inequality
violation
\begin{align}
{\cal S}\left(\rho_{A B C}\right)=\max \left\{{\cal S}_{A B}, {\cal S}_{A C},  {\cal S}_{B C}\right\}.
\label{Eq.Sabc}
\end{align}

\section{GENUINE TRIPARTITE ENTANGLEMENT VERSUS FIRST-ORDER COHERENCE\label{sec3}}
In this section, we present the intrinsic relations between the genuine tripartite entanglement and the first-order coherence
for three-qubit pure states. In particular, we show that there exists an exact functional relation between GGM and GMC. Moreover,
the tradeoff relations between genuine tripartite entanglement, such as GMC and concurrence fill, and first-order coherence in
the context of an arbitrary three-qubit pure state are established.
These correspondence relations may deepen the understanding of the interconversions among different measures of nonclassical
correlations in the framework of QRTs.

In order to reveal these relations in a more explicit manner, here we introduce three boundary states with a single parameter. The first one is the generalized GHZ state, which can exhibit
maximum first-order coherence value for a fixed amount of genuine tripartite entanglement, i.e.,
\begin{align}
{\left| \psi  \right\rangle _\alpha } = \cos \alpha \left| {i,j,k} \right\rangle  + \sin \alpha {\rm{ }}\left| {\bar i,\bar j,\bar k} \right\rangle ,
\label{Eq.A14}
\end{align}
where $i,j,k \in \{ 0,1\} $ and the superscript "$-$" means taking the opposite value.
Since their performance are equivalent, we take the following states as an example in the calculation
\begin{align}
{\left| \psi  \right\rangle _\alpha } = \cos \alpha \left| {000} \right\rangle  + \sin \alpha \left| {111} \right\rangle .
\label{Eq.A14}
\end{align}

The second boundary state is a single parameter family of three-qubit pure state with
\begin{align}
|\psi\rangle_{m}=\frac{|000\rangle+m(|010\rangle+|101\rangle)+|111\rangle}{\sqrt{2+2 m^{2}}},
\label{Eq.A15}
\end{align}
where $m \in [0,1]$. For $m \in [0,1)$, the state belongs to the GHZ class, and the state belongs to the W class when $m = 1$.
Interestingly, this class of state is also regarded as the
 the maximally steering inequality violating states \cite{PhysRevA.105.022425}, maximally Bell-inequality violating states \cite{PhysRevA.94.052126} and the
maximally dense-coding-capable states \cite{PhysRevA.87.032336}.\par

The third one is a single parameter family of separable three-qubit pure state, which is located in the upper boundary of the relation between the maximum steering inequality violation and first-order coherence.
It is given by
\begin{align}
{\left| \psi  \right\rangle _\theta } = \left| i \right\rangle \left( {\cos \theta \left| {j,k} \right\rangle  + \sin \theta \left| {\bar j,\bar k} \right\rangle } \right)
\end{align}
where $i,j,k \in \{ 0,1\}$ and $\left| i  \right\rangle$ also can represent the second or third qubit.
We choose the following states as an example,
\begin{align}
{\left| \psi  \right\rangle _\theta } = \cos \theta \left| {001} \right\rangle  + \sin \theta \left| {100} \right\rangle  ,
\label{Eq.A17}
\end{align}
Note that the above three boundary states always form a trilateral region in the investigation of unification of different measures of quantum resources,
in which all the three-qubit pure states will be included.

\subsection{GMC versus GGM}
The exact relation between GMC and GGM for three-qubit pure states is derived in
this section.\par
{\it Theorem 1.\,}
For a three-qubit pure state $|\psi \rangle$, GGM and GMC satisfy the following relation
 \begin{align}
(2{\cal G}(|\psi \rangle ) - 1{)^2} + {\cal C }(|\psi \rangle {)^2} = 1,
\label{Eq.1}
\end{align}
where $0 \le {\cal G}(|\psi \rangle ) \le 1/2$ and $0 \le {\cal C}(|\psi \rangle ) \le 1$.
\par\begin{proof}
For a three-qubit pure state $|\psi \rangle$, the GGM is given by:
\begin{align}
{\cal G}(|\psi \rangle ) = 1 - \max \{ {\lambda _1},{\lambda _3},{\lambda _5}\}  = min\{ {\lambda _2},{\lambda _4},{\lambda _6}\} ,
\label{Eq.GGM3}
\end{align}
where ${\lambda _1}$, ${\lambda _3}$, and ${\lambda _5}$ are the bigger eigenvalues of the reduced density
matrices ${\rho _A}$, ${\rho _B}$, and ${\rho _C}$, respectively, and ${\lambda _2}$, ${\lambda _4}$, and ${\lambda _6}$ are the smaller ones.
The second equation is obtained from the trace condition of reduced density
matrices
\begin{align}
{\lambda _1} + {\lambda _2} = 1,\quad {\lambda _3} + {\lambda _4} = 1,\quad {\lambda _5} + {\lambda _6} = 1.
\label{Eq.con1}
\end{align}
If we assume that
\begin{align}
{\lambda _2}  \le  {\lambda _4},\quad {\lambda _2} \le {\lambda _6},
\label{Eq.con2}
\end{align}
then one can get the GGM of the state $|\psi \rangle$ as
\begin{align}
{\cal G}(|\psi \rangle ) = {\lambda _2},
\label{Eq.GGM4}
\end{align}

The GMC of three-qubit pure states is given by:
\begin{align}
{\cal C}(|\psi \rangle ) = \mathop {\min }\limits_i \sqrt {2\left[ {1 - {\mathop{\rm Tr}\nolimits} \left( {\rho _i^2} \right)} \right]}
\label{Eq.GMC3},
\end{align}
where $i \in \{ A,B,C\} $ and ${\mathop{\rm Tr}\nolimits} \left( {\rho _i^2} \right)$ is the purity of the reduced density
matrices. It can be calculated as
\begin{align}
{\rm{Tr}}\left( {\rho _A^2} \right) \!= \!\lambda _1^2\! +\! \lambda _2^2,\,\,{\rm{Tr}}\left( {\rho _B^2} \right) \!= \!\lambda _3^2 \!+\! \lambda _4^2,\,\,{\rm{Tr}}\left( {\rho _C^2} \right) \!= \!\lambda _5^2 \!+\! \lambda _6^2.
\label{Eq.purity}
\end{align}
From the Eqs. (\ref{Eq.con1}), (\ref{Eq.con2}), and (\ref{Eq.purity}), one can show that (see Appendix \hyperlink{A}{A})
\begin{align}
{\rm{Tr}}\left( {\rho _A^2} \right) \ge{\rm{  Tr}}\left( {\rho _B^2} \right),\quad {\rm{Tr}}\left( {\rho _A^2} \right)\ge{\rm{  Tr}}\left( {\rho _C^2} \right).
\label{Eq.cond1}
\end{align}
This gives
 \begin{align}
 {\cal C}(|\psi \rangle ) = \sqrt {2\left[ {1 - {\rm{Tr}}\left( {\rho _A^2} \right)} \right]} .
\label{Eq.GMC4}
\end{align}
From the Eqs. (\ref{Eq.con1}), (\ref{Eq.GGM4}), (\ref{Eq.purity}), and (\ref{Eq.GMC4}), one can finally obtain the relation between GGM and GMC as the Eq. (\ref{Eq.1}).
The above relation also holds if we assume ${\lambda _4}$ or ${\lambda _6}$ is the smallest one among the eigenvalues ${\lambda _2}$, ${\lambda _4}$, and ${\lambda _6}$.
\end{proof}

\begin{figure}
\centering
\includegraphics[width=8.6cm]{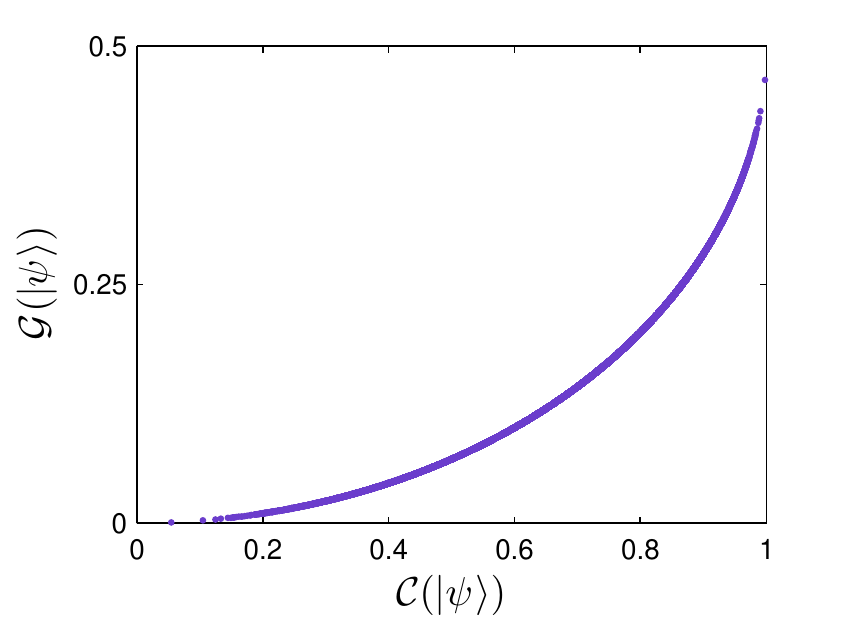}
\caption{Exact functional relation between the GGM, ${\cal G}(|\psi \rangle )$, and the GMC, ${\cal C}(|\psi \rangle )$, for ${10^5}$ Haar randomly generated three-qubit pure states. The $x$ and $y$ axes are dimensionless.}
\label{f1}
\end{figure}

In Fig. \ref{f1}, we plot the exact functional relation between the GGM and GMC for ${10^5}$ Haar randomly generated three-qubit pure states \cite{bengtsson_zyczkowski_2017,*Zyczkowski_1994}.
The results shows that the GGM is one quarter of elliptic curve with respect to GMC, whose center point is located at $(0,0.5)$. The minor axis of the ellipse lies at the longitudinal axis with value being 1
and the major axis is 2.

\subsection{First-order coherence versus GMC}
The tradeoff relation between first-order coherence and GMC for three-qubit pure states is derived in
this section.\par
{\it Theorem 2.\,}
If a three-qubit pure state $|\psi \rangle$ has
the same value of GMC with boundary states $|\psi \rangle _{\alpha}$ and $|\psi \rangle _{m}$, the first-order coherence
of these three states satisfy the ordering ${\cal D}{(|\psi \rangle _m}) \le {\cal D}(|\psi \rangle ) \le {\cal D}{(|\psi \rangle _{\alpha}})$.
And the tradeoff relation of GMC and first-order coherence is given by
\begin{align}
\left\{ \begin{array}{l}\vspace{1.5ex}
{\cal C}(|\psi \rangle {)^2} + {\cal D}(|\psi \rangle {)^2} \le 1 \\
{\cal C}(|\psi \rangle {)^2} + 3{\cal D}(|\psi \rangle {)^2} \ge 1
\end{array} \right.
\label{Eq.2}
\end{align}
\par\begin{proof}
For the state $|\psi \rangle$, from Eqs. (\ref{Eq.D1}) and (\ref{Eq.Dabc}), the square of its first-order coherence can be obtained as
\begin{align}
{\cal D}(|\psi \rangle {)^2} = \frac{2}{3}\left[ {Tr(\rho _A^2) + Tr(\rho _B^2) + Tr(\rho _C^2)} \right] - 1.
\label{Eq.35a}
\end{align}
Assume that
\begin{align}
{\rm{Tr}}\left( {\rho _A^2} \right) \ge {\rm{  Tr}}\left( {\rho _B^2} \right),\quad {\rm{Tr}}\left( {\rho _A^2} \right) \ge {\rm{ Tr}}\left( {\rho _C^2} \right),
\end{align}
we can obtain
\begin{align}
 {\cal C}(|\psi \rangle {)^2} = 2\left[ {1 - {\rm{Tr}}\left( {\rho _A^2} \right)} \right] ,
\label{Eq.37a}
\end{align}
From this inequality,
\begin{align}
 {\rm{Tr}}\left( {\rho _B^2} \right) + {\rm{Tr}}\left( {\rho _C^2} \right) \le 2{\rm{Tr}}\left( {\rho _A^2} \right),
\label{Eq.38a}
\end{align}
one can see (see Appendix \hyperlink{B 1}{B 1})
\begin{align}
 2\left[ {1 \!- \!{\rm{Tr}}\left( {\rho _A^2} \right)} \right] \!+\! \frac{2}{3}\left[ {Tr(\rho _A^2) \!+\! Tr(\rho _B^2) \!+\! Tr(\rho _C^2)} \right] \!-\! 1 \le 1.
\label{Eq.39a}
\end{align}
Therefore, for the state $|\psi \rangle$, substituting the Eqs. (\ref{Eq.35a}) and (\ref{Eq.37a}) into Eq. (\ref{Eq.39a}), we get the upper boundary of the relation between GMC and first-order coherence
\begin{align}
 {\cal C}(|\psi \rangle {)^2} + {\cal D}(|\psi \rangle {)^2} \le 1.
\label{Eq.40a}
\end{align}
Based on the fact that $Tr(\rho _i^2)\geq \frac{1}{2}$, where $i \in \{ A,B,C\} $, we have
\begin{align}
 {\rm{Tr}}\left( {\rho _B^2} \right) + {\rm{Tr}}\left( {\rho _C^2} \right) \ge 1.
\label{Eq.41a}
\end{align}
By this inequality, one can prove that (see Appendix \hyperlink{B 2}{B 2})
\begin{align}
2\left[ {1\! -\! {\rm{Tr}}\left( {\rho _A^2} \right)} \right] \!+\! 2\left[ {Tr(\rho _A^2) \!+\! Tr(\rho _B^2) \!+\! Tr(\rho _C^2)} \right] \!- \!3 \ge 1.
\label{Eq.42a}
\end{align}
Similarly, we obtain the lower boundary of the relation between GMC and first-order coherence as
\begin{align}
{\cal C}(|\psi \rangle {)^2} + 3{\cal D}(|\psi \rangle {)^2} \ge 1.
\label{Eq.43a}
\end{align}
Moreover, the relations are also valid if we assume ${\rm{Tr}}\left( {\rho _B^2} \right)$ or ${\rm{Tr}}\left( {\rho _C^2} \right)$ is the largest
one among the three purities of subsystems $\rho _A$, $\rho _B$, and $\rho _C$.

The GMC and first-order coherence of state $|\psi \rangle _{\alpha}$ and $|\psi \rangle _{m}$, from Eqs. (\ref{Eq.Dabc}) and (\ref{Eq.GMC3}), are given by
\begin{align}
{\cal C}{(|\psi \rangle _\alpha }) = \sqrt {2(1 - \cos^4\alpha  - \sin^4\alpha )} ,
\label{Eq.GMC5}
\end{align}
\begin{align}
{\cal D}{(|\psi \rangle _\alpha }) = \left| {\cos 2\alpha } \right| ,
\label{Eq.Dabc2}
\end{align}
\begin{align}
{\cal C}{(|\psi \rangle _m}) = \frac{{1 - {m^2}}}{{1 + {m^2}}},
\label{Eq.GMC6}
\end{align}
\begin{align}
{\cal D}{(|\psi \rangle _m}) = \frac{{2m}}{{\sqrt 3 (1 + {m^2})}},
\label{Eq.Dabc3}
\end{align}
respectively. One can find that
\begin{align}
{\cal C}{(|\psi \rangle _\alpha }{)^2} + {\cal D}{(|\psi \rangle _\alpha }{)^2} = 1,
\label{Eq.21}
\end{align}
\begin{align}
{\cal C}{(|\psi \rangle _m}{)^2} + 3{\cal D}{(|\psi \rangle _m}{)^2} = 1,
\label{Eq.22}
\end{align}
which imply that states $|\psi \rangle _{\alpha}$ and $|\psi \rangle _{m}$ are the upper and lower boundary states.
\end{proof}

\begin{figure}
\centering
\includegraphics[width=8.6cm]{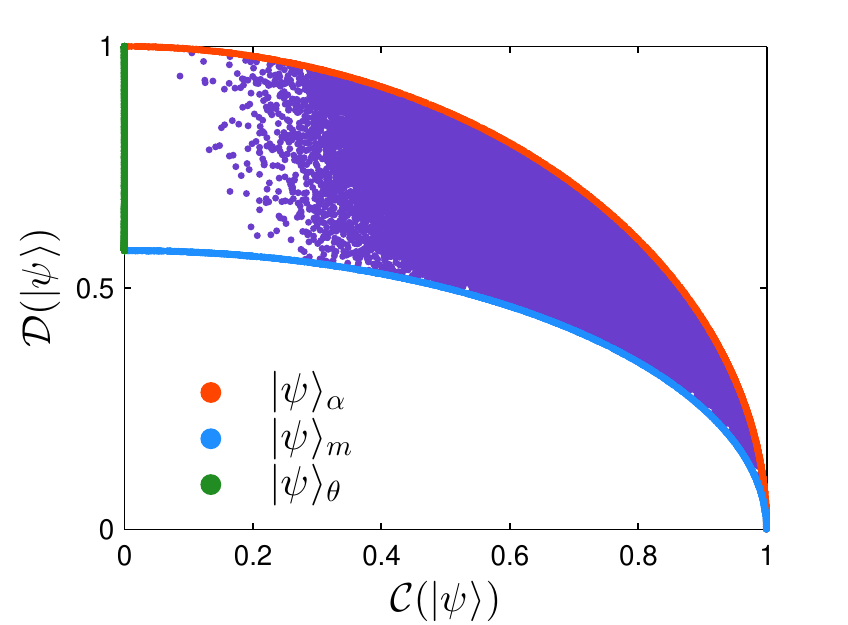}
\caption{(Color online)  Tradeoff relation between the first-order coherence, ${\cal D}(|\psi \rangle )$, and the GMC, ${\cal C}(|\psi \rangle )$, for ${10^5}$ Haar randomly generated three-qubit pure states. The orange red line is the upper boundary with the sate $|\psi \rangle _\alpha$, the state $|\psi \rangle _m$ lies at the Dodger blue lower boundary line, and the forest green line represents the state $|\psi \rangle _\theta$, which lies at the $y$-axis. The $x$ and $y$ axes are dimensionless.}
\label{f2}
\end{figure}

In Fig. \ref{f2}, we plot how the first-order coherence changes with respect to the GMC for ${10^5}$ Haar randomly generated three-qubit pure states. The orange red upper boundary line donates the state $|\psi \rangle _\alpha$ satisfying the relation of Eq. (\ref{Eq.21}) between first-order coherence and GMC. The Dodger blue lower boundary line shows that the two quantum resources of the state $|\psi \rangle _m$ fulfilling the
relation of Eq. (\ref{Eq.22}). The ${10^5}$ Haar randomly generated three-qubit pure states are included in the trilateral region formed by states $|\psi \rangle _\alpha$,  $|\psi \rangle _m$, and $|\psi \rangle _\theta$, meaning their first-order coherence and GMC obey the inequalities in Eq. (\ref{Eq.2}). Moreover, we find the first-order coherence increases (decrease) with the decrease (increase) of the GMC, shows a tradeoff.

\subsection{First-order coherence versus concurrence fill}
The tradeoff relation between first-order coherence and concurrence fill for three-qubit pure states is derived in
this section.\par
{\it Theorem 3.\,}
If a three-qubit pure state $|\psi \rangle$ has
the same value of concurrence fill with state $|\psi \rangle _{\alpha}$ and $|\psi \rangle _m$, the first-order coherence
of these three states satisfy the ordering ${\cal D}{(|\psi \rangle _m}) \le {\cal D}(|\psi \rangle ) \le {\cal D}{(|\psi \rangle _{\alpha}})$.
And the tradeoff relation of concurrence fill and first-order coherence is given by
\begin{align}
\left\{ \begin{array}{l}\vspace{1.5ex}
\!{\cal F}(|\psi \rangle ) + {\cal D}(|\psi \rangle {)^2} \le 1 ,\\
\!{\cal F}(|\psi \rangle {)^4}\! + \!{(3{\cal D}(|\psi \rangle )^2}\! -\! 1{)^2}{(3{\cal D}(|\psi \rangle )^4} \!-\! 2{\cal D}(|\psi \rangle {)^2} \!-\! 1) \!\ge\! 0,
\end{array} \right.
\label{Eq.3}
\end{align}
where $ 0 \le {\cal D}(|\psi \rangle ) \le 1/\sqrt 3 $ for the second inequality.
\par\begin{proof}
For the state $|\psi \rangle$, substituting the Eq. (\ref{Eq.purity}) into the Eq. (\ref{Eq.35a}), its first-order coherence can be written as
\begin{align}
{\cal D}(|\psi \rangle ){^2} = \frac{2}{3}\left( {\lambda _1^2 + \lambda _2^2 + \lambda _3^2 + \lambda _4^2 + \lambda _5^2 + \lambda _6^2} \right) - 1.
\label{Eq.D31}
\end{align}
From the Eqs. (\ref{Eq.Q}) and (\ref{Eq.C}), one can get
\begin{align}
Q = 2\left( {{\lambda _1}{\lambda _2} + {\lambda _3}{\lambda _4} + {\lambda _5}{\lambda _6}} \right).
\label{Eq.Q31}
\end{align}
Then one can obtain the relation between ${\cal D}(|\psi \rangle )$ and Q
\begin{align}
{\cal D}(|\psi \rangle {)^2} = 1 - \frac{2}{3}Q.
\label{Eq.51a}
\end{align}
For simplicity, we define ${C^2_{A(BC)}}$, ${C^2_{B(AC)}}$, and ${C^2_{C(AB)}}$
as $a$, $b$, and $c$, respectively. By the mean value inequality, we have
\begin{align}
(Q - a)(Q - b)(Q - c) \le {\left( {\frac{Q}{3}} \right)^3}.
\label{Eq.52a}
\end{align}
Note that the summation of three terms in each side of the inequality is equal to $Q$. As a result, one can find that (see Appendix \hyperlink{C 1}{C 1})
\begin{align}
\begin{split}
 {\left[ {\frac{{16}}{3}Q\left( {Q - a} \right)\left( {Q - b} \right)\left( {Q - c} \right)} \right]^{1/4}} + 1 - \frac{2}{3}Q \le 1,
 \end{split}
\label{Eq.53a}
\end{align}
Substituting the Eqs. (\ref{Eq.F123}) and (\ref{Eq.51a}) into the Eq. (\ref{Eq.53a}), we have
\begin{align}
{\cal F}(|\psi \rangle ) + {\cal D}(|\psi \rangle {)^2} \le 1.
\label{Eq.54a}
\end{align}
On the other hand, from the Eq. (\ref{Eq.51a}) and the relation $ 0 \le {\cal D}(|\psi \rangle ) \le 1/\sqrt 3$, one can obtain
$1 \le Q \le 3/2$. Since $0 \le a,b,c \le 1$, we get that $Q - 1 \le Q - a,Q - b,Q - c \le 2 - Q$.
Thus, using the mean value inequality, we have
\begin{align}
(2 - Q){(Q - 1)^2} \le (Q - a)(Q - b)(Q - c).
\label{Eq.55a}
\end{align}
Consequently, one can see (see Appendix \hyperlink{C 2}{C 2})
\begin{align}
\begin{split}
 &\frac{{16}}{3}Q\left( {Q - a} \right)\left( {Q - b} \right)\left( {Q - c} \right)\\
 &\! + \!{\left[ {3\left( {1 \!-\! \frac{2}{3}Q} \right) \!-\! 1} \right]^2}\left[ {3{{\left( {1\! -\! \frac{2}{3}Q} \right)}^2} \!- \!2\left( {1 \!- \!\frac{2}{3}Q} \right)\! - \!1} \right] \!\ge\! 0.
 \end{split}
\label{Eq.56a}
\end{align}
In a similar way, we have
\begin{align}
\!{\cal F}(|\psi \rangle {)^4}\! + \!{(3{\cal D}(|\psi \rangle )^2}\! -\! 1{)^2}{(3{\cal D}(|\psi \rangle )^4} \!-\! 2{\cal D}(|\psi \rangle {)^2} \!-\! 1) \!\ge\! 0.
\label{Eq.57a}
\end{align}
\indent The concurrence fill of state $|\psi \rangle _{\alpha}$ and $|\psi \rangle _{m}$, from Eq. (\ref{Eq.F123}), are given by
\begin{align}
{\cal F}{(|\psi \rangle _\alpha }) = {\sin ^2}\left( {2\alpha } \right),
\label{Eq.F31}
\end{align}
\begin{align}
{\cal F}(|\psi \rangle _m)\!\!=\!\! \frac{{\left( {1 - {m^2}} \right)\!{{\left[{\left( {1 + 6{m^2} \!+\! {m^4}} \right)\!\left( {3 \!+\! 2{m^2}\! + \!3{m^4}} \right)} \right]}^{1/4}}}}{{{3^{1/4}}{{\left( {1 + {m^2}} \right)}^2}}}.
\label{Eq.F32}
\end{align}
Together with the Eqs. (\ref{Eq.Dabc2}) and (\ref{Eq.Dabc3}), one can obtain the following relations
\begin{align}
{\cal F}{(|\psi \rangle _\alpha }) + {\cal D}{(|\psi \rangle _\alpha }{)^2} = 1,
\label{Eq.F33}
\end{align}
\begin{align}
\begin{split}
 {\cal F}{(|\psi \rangle _m}{)^4}\! &+\! {(3 {\cal D} {(|\psi \rangle _m})^2} \! - \! 1{)^2} \\
 &{\rm{              }} \times {(3 {\cal D} {(|\psi \rangle _m})^4} \!- \!2{\cal D}{(|\psi \rangle _m}{)^2}\! -\! 1) \!\!=\!\! 0,
\label{Eq.F34}
\end{split}
\end{align}
which imply that states $|\psi \rangle _{\alpha}$ and $|\psi \rangle _{m}$, respectively, are the upper and lower boundary states.
\end{proof}

\begin{figure}
\centering
\includegraphics[width=8.6cm]{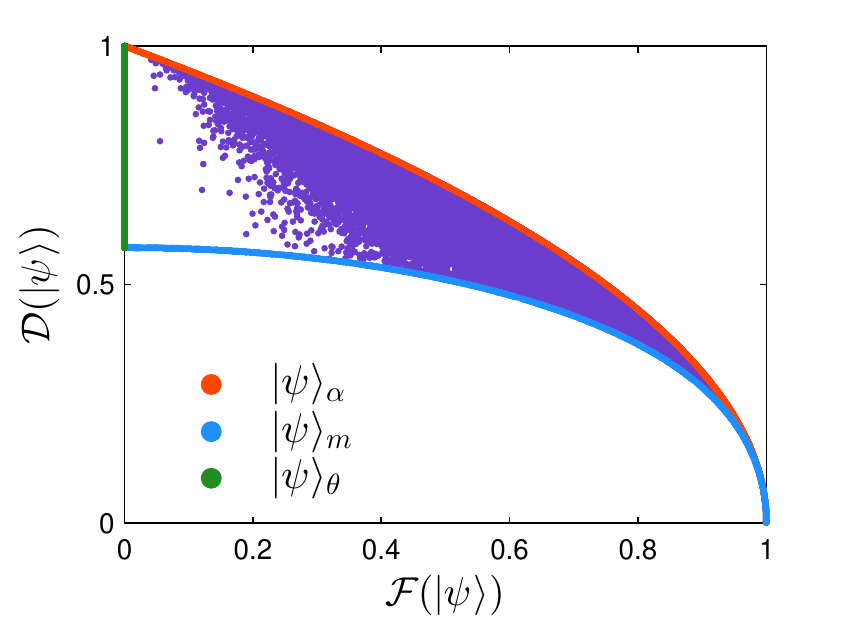}
\caption{(Color online)  Tradeoff relation between the first-order coherence, ${\cal D}(|\psi \rangle )$, and the concurrence fill, ${\cal F}(|\psi \rangle )$, for ${10^5}$ Haar randomly generated three-qubit pure states. The orange red line is the upper boundary with the sate $|\psi \rangle _\alpha$, the state $|\psi \rangle _m$ lies at the Dodger blue lower boundary line, and the forest green line represents the state $|\psi \rangle _\theta$, which lies at the $y$-axis. The $x$ and $y$ axes are dimensionless.}
\label{f3}
\end{figure}

In Fig. \ref{f3}, the relation between the first-order coherence and the concurrence fill is plotted, for ${10^5}$ Haar randomly generated three-qubit pure states. We can find that the states $|\psi \rangle _\alpha$ and $|\psi \rangle _m$ donate the upper (orange red line) and lower (Dodger blue line) boundary lines, which means that their first-order coherence and concurrence fill satisfy the relations in Eqs. (\ref{Eq.F33}) and  (\ref{Eq.F34}), respectively. Together with the state $|\psi \rangle _\theta$ in the y-axis, they form a trilateral region, which includes all the three-qubit pure states. Also, its shows that there exists a trade-off relation between the the first-order coherence and concurrence fill for arbitrary three-qubit pure states.

\section{THE MAXIMUM STEERING INEQUALITY VIOLATION VERSUS CONCURRENCE FILL\label{sec4}}
 Quantum steering describes an important trait of the quantum world that one system can immediately affect another one by local measurements. The concurrence fill is introduced as a good triangle measure of tripartite entanglement in 2021, which can detect genuine three-qubit entanglement faithfully. Here, our aim is to
study the relation between the maximum steering inequality violation and concurrence fill.   \par
{\it Theorem 4.\,}
If a three-qubit pure state $|\psi \rangle$ has
the same value of concurrence fill with the state $|\psi \rangle _m$, the maximum steering inequality violation
of these two states satisfy the ordering $  {\cal S}(|\psi \rangle ) \le {\cal S}{(|\psi \rangle _m})$.
And the tradeoff relation of the maximum steering inequality violation and concurrence fill is given by
\begin{align}
48{\cal F}(|\psi \rangle {)^4} + ({\cal S}(|\psi \rangle ) - 3{)^2}({\cal S}(|\psi \rangle ) + 1)({\cal S}(|\psi \rangle ) - 7) \le 0.
\label{Eq.4}
\end{align}
\par\begin{proof}
Any three-qubit state ${\rho _{ABC}}$ can be written as
\begin{align}
\begin{split}
\rho_{A B C}=& \frac{1}{8} \Bigg[ \mathbb{I} \otimes \mathbb{I} \otimes \mathbb{I}+\vec{A} \cdot \vec{\sigma} \otimes \mathbb{I} \otimes \mathbb{I}+\mathbb{I} \otimes \vec{B} \cdot \vec{\sigma} \otimes \mathbb{I}\\
&+\mathbb{I} \otimes \mathbb{I} \otimes \vec{C} \cdot \vec{\sigma}+\sum_{i j} t_{i j}^{A B} \sigma_{i} \otimes \sigma_{j} \otimes \mathbb{I} \\
&+\sum_{i k} t_{i k}^{A C} \sigma_{i} \otimes \mathbb{I} \otimes \sigma_{k}+\sum_{j k} t_{j k}^{B C} \mathbb{I} \otimes \sigma_{j} \otimes \sigma_{k} \\
&\left.+\sum_{i j k} t_{i j k}^{A B C} \sigma_{i} \otimes \sigma_{j} \otimes \sigma_{k} \right].
\end{split}
\label{Eq.4rho}
\end{align}
This gives
\begin{align}
\operatorname{tr}\left(\rho_{A}^{2}\right)=\frac{1+\vec{A}^{2}}{2}, \quad \operatorname{Tr}\left(\rho_{B C}^{2}\right)=\frac{1}{4}\left(1+\vec{B}^{2}+\vec{C}^{2}+{\cal S}_{B C}\right).
\label{Eq.42}
\end{align}
Similarly, we have
\begin{align}
\operatorname{tr}\left(\rho_{B}^{2}\right)=\frac{1+\vec{B}^{2}}{2}, \quad \operatorname{Tr}\left(\rho_{A C}^{2}\right)=\frac{1}{4}\left(1+\vec{A}^{2}+\vec{C}^{2}+{\cal S}_{A C}\right)\nonumber,
\end{align}
\begin{align}
\operatorname{tr}\left(\rho_{C}^{2}\right)=\frac{1+\vec{C}^{2}}{2}, \quad \operatorname{Tr}\left(\rho_{A B}^{2}\right)=\frac{1}{4}\left(1+\vec{A}^{2}+\vec{B}^{2}+{\cal S}_{A B}\right).
\label{Eq.43}
\end{align}
If $\rho_{ABC}$ is a pure state, by the Schimdt decomposition, we have $Tr(\rho _i^2) = Tr(\rho _{jk}^2)$
for i $\ne j \ne k,\,\,i,j,k \in \{ A,B,C\} $. From Eqs.(\ref{Eq.42}) and (\ref{Eq.43}),
we can write ${\cal S}_{AB}$ as the function of the purities of the subsystems
 \begin{align}
{{\cal S}_{AB}} = 4Tr(\rho _C^2) - 2Tr(\rho _A^2) - 2Tr(\rho _B^2) + 1.
\label{Eq.60a}
\end{align}
Combining the above equation and Eqs. (\ref{Eq.C}), (\ref{Eq.con1}) and (\ref{Eq.purity}), one can obtain that (see Appendix \hyperlink{D 1}{D 1})
 \begin{align}
{{\cal S}_{AB}} = a + b - 2c + 1.
\label{Eq.S43}
\end{align}
Similarly, we get
 \begin{align}
{{\cal S}_{AC}} = a + c - 2b + 1\nonumber,\\
{{\cal S}_{BC}} = b + c - 2a + 1.
\label{Eq.S44}
\end{align}
Assume that the bipartite steering of the subsystem ${S_{AB}}$ is the largest one among ${S_{AB}}$, ${S_{AC}}$, and ${S_{BC}}$, i.e., ${S_{AB}} \ge {S_{AC}},\,{S_{AB}} \ge {S_{BC}}$,
Thus, we get the maximum steering inequality violation
 \begin{align}
{\cal S}(|\psi \rangle)  = {{\cal S}_{AB}},
\label{Eq.S45}
\end{align}
From the Eqs. (\ref{Eq.S43}) and (\ref{Eq.S44}), we have $a\ge c$, $b\ge c$, and $0 \le a + b - 2c \le 2$. By these constraints, one can show that (see Appendix \hyperlink{D 2}{D 2})
\begin{align}
 \begin{split}
4(Q - a) + 4(Q - b) \le 2\left[ {2 - (a + b - 2c)} \right],
\label{Eq.68b}
\end{split}
\end{align}
Using the mean value inequality, we get
\begin{align}
 \begin{split}
4(Q - a) \cdot 4(Q - b) \le {\left[ {2 - (a + b - 2c)} \right]^2},
\label{Eq.69b}
\end{split}
\end{align}
Similar to Eq. (\ref{Eq.68b}), one can obtain
 \begin{align}
 \begin{split}
4(Q - c)\le&2 + a + b - 2c,\\
4Q\le&6 - (a + b - 2c).
\label{Eq.S45}
\end{split}
\end{align}
As a consequence, we have
\begin{align}
\begin{split}
4(Q - a)\cdot4(Q - b)\cdot4(Q - c)\cdot4Q \le &{ [2 - (a + b - 2c)]^2}\\
\times\left( {2 + a + b - 2c} \right)&[6 - (a + b - 2c)].
\end{split}
\label{Eq.71c}
\end{align}
From this inequality, one can see that (see Appendix \hyperlink{D 3}{D 3})
\begin{align}
\begin{split}
&48 \times \frac{{16}}{3}Q\left( {Q - a} \right)\left( {Q - b} \right)\left( {Q - c} \right)\\
& \!+\! \!{\left( {a \!+\! b \!-\! 2c \!+\! 1\! - \!3} \right)^2}\!\!\left( {a \!+ \!b \!-\! 2c \!+\! 1\! +\! 1} \right)\!\left( {a \!+\! b \!-\! 2c \!+\!1 \!-\! 7} \right) \!\!\le\!\! 0.
\end{split}
\label{Eq.70b}
\end{align}
Finally, substituting the Eqs. (\ref{Eq.F123}) and (\ref{Eq.S45}) into the Eq. (\ref{Eq.70b}), we obtain the tradeoff relation
of the maximum steering inequality violation and concurrence fill for three-qubit pure states as
\begin{align}
48{\cal F}(|\psi \rangle {)^4} + ({\cal S}(|\psi \rangle ) - 3{)^2}({\cal S}(|\psi \rangle ) + 1)({\cal S}(|\psi \rangle ) - 7) \le 0.
\label{Eq.71b}
\end{align}
The tradeoff relation also holds for the situations that the bipartite steering ${S_{AC}}$ or ${S_{BC}}$ is the largest one among ${S_{AB}}$, ${S_{AC}}$, and ${S_{BC}}$.

The maximum steering inequality violation of the state $|\psi \rangle _{m}$, from Eq. (\ref{Eq.Sabc}), can be calculated as
\begin{align}
{\cal S}{(|\psi \rangle _m}) = \frac{{1 + 10{m^2} + {m^4}}}{{{{(1 + {m^2})}^2}}}.
\label{Eq.S41}
\end{align}
Together with Eq. (\ref{Eq.F32}), one can obtain
\begin{align}
48{\cal F}{(|\psi \rangle _m}\!{)^4} \!+\! ({\cal S}{(|\psi \rangle _m}\!) \!-\! 3{)^2}({\cal S}{(|\psi \rangle _m}\!) \!+\! 1)({\cal S}{(|\psi \rangle _m}\!) \!-\! 7) \!\!=\!\! 0.
\label{Eq.41}
\end{align}
which imply that state $|\psi \rangle _{m}$ is the upper boundary states.
\end{proof}

\begin{figure}
\centering
\includegraphics[width=8.6cm]{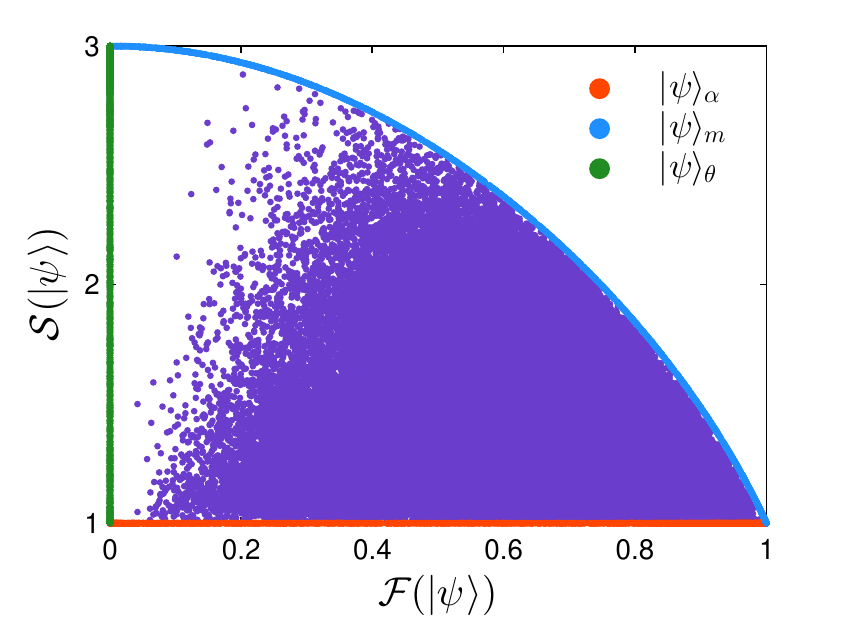}
\caption{(Color online) Tradeoff relation between the maximum steering inequality violation, ${\cal S}(|\psi \rangle )$, and the concurrence fill, ${\cal F}(|\psi \rangle )$, for ${10^5}$ Haar randomly generated three-qubit pure states. The Dodger blue line is the upper boundary representing the sate $|\psi \rangle _m$, the states $|\psi \rangle _\alpha$ and $|\psi \rangle _\theta$ lie at $x$ and $y$ axes, respectively. The $x$ and $y$ axes are dimensionless.}
\label{f4}
\end{figure}

In Fig. \ref{f4}, we plot the relation between the maximum steering inequality violation and the concurrence fill for ${10^5}$ Haar randomly generated three-qubit pure states.
One can see that the state $|\psi \rangle _m$ located at the upper boundary line (Dodger blue line), suggesting that its maximum steering inequality violation and concurrence fill satisfy the Eq. (\ref{Eq.41}). In particular, the states $|\psi \rangle _\alpha$ and $|\psi \rangle _\theta$ lie at $x$ and $y$ axes, respectively. The purple dots donate ${10^5}$ Haar randomly generated three-qubit pure states.

\section{THE MAXIMUM STEERING INEQUALITY VIOLATION VERSUS FIRST-ORDER COHERENCE\label{sec5}}
The close relation between the maximum steering inequality violation and first-order coherence for three-qubit pure states is derived in
this section.\par
{\it Theorem 5.\,}
If a three-qubit pure state $|\psi \rangle$ has
the same value of first-order coherence with state
 $|\psi \rangle _m$ or $|\psi \rangle _{\theta}$,
the maximum steering inequality violation
of these three states satisfy the ordering ${\cal S}(|\psi \rangle )
\le {\cal S}{(|\psi \rangle _m})$ or ${\cal S}(|\psi \rangle ) \le {\cal S}{(|\psi \rangle _{\theta}})$.
And the relation between the maximum steering inequality violation and first-order coherence is given by
\begin{align}
\left\{ {\begin{array}{*{20}{l}}\vspace{1.5ex}
{{\cal S}(|\psi \rangle ) - 6{\cal D}(|\psi \rangle {)^2} \le 1,\quad 0 \le {\cal D}(|\psi \rangle ) < \frac{1}{{\sqrt 3 }}}\\
{{\cal S}(|\psi \rangle ) + 3{\cal D}(|\psi \rangle {)^2} \le 4,\quad \frac{1}{{\sqrt 3 }} \le {\cal D}(|\psi \rangle ) \le 1}
\end{array}} \right.
\label{Eq.5}
\end{align}
\par\begin{proof}
 In the following, we assume that ${\cal S}{(|\psi \rangle })={\cal S}_{AB}$.
Similar to Eq. (\ref{Eq.41a}), we have
\begin{align}
Tr(\rho _A^2) + Tr(\rho _B^2) \ge 1.
\label{Eq.71a}
\end{align}
By this inequality, one can show that (see Appendix \ref{app:E1})
\begin{align}
\begin{split}
4Tr(\rho _C^2) &- 2Tr(\rho _A^2) -  2Tr(\rho _B^2) + 1 \\
&- 6\left[ {\frac{2}{3}\left( {Tr(\rho _A^2) +  Tr(\rho _B^2) + Tr(\rho _C^2)} \right) - 1} \right] \le 1,
\end{split}
\label{Eq.72a}
\end{align}
From Eqs. (\ref{Eq.35a}) and (\ref{Eq.60a}), one can obtain
\begin{align}
 {\cal S}(|\psi \rangle ) - 6{\cal D}(|\psi \rangle {)^2} \le 1.
\label{Eq.73a}
\end{align}

On the other hand, from inequality $Tr(\rho _C^2) \le 1$, one can find that (see Appendix \ref{app:E2})
\begin{align}
\begin{split}
4Tr(\rho _C^2) &- 2Tr(\rho _A^2) - 2Tr(\rho _B^2) + 1 \\
&+ 3\left[ {\frac{2}{3}\left( {Tr(\rho _A^2) + Tr(\rho _B^2) + Tr(\rho _C^2)} \right) - 1} \right] \le 4,
\end{split}
\label{Eq.75a}
\end{align}
then we have
\begin{align}
 {\cal S}(|\psi \rangle ) + 3{\cal D}(|\psi \rangle {)^2} \le 4.
\label{Eq.76a}
\end{align}
The close relation in Eq. (\ref{Eq.5}) holds when ${\cal S}{(|\psi \rangle })={\cal S}_{AC}$ or ${\cal S}{(|\psi \rangle })={\cal S}_{BC}$.

The maximum steering inequality violation and first-order coherence of the state $|\psi \rangle _{\theta}$, from Eqs. (\ref{Eq.Dabc}) and (\ref{Eq.Sabc}), are
\begin{align}
{\cal S}{(|\psi \rangle _\theta }) = 2 - \cos(4\theta ),\\
{\cal D}{(|\psi \rangle _\theta }) = \sqrt {\frac{{2 + \cos(4\theta )}}{3}} .
\label{Eq.SD51}
\end{align}
Together with Eqs. (\ref{Eq.Dabc3}) and (\ref{Eq.S41}), one can obtain
\begin{align}
{\cal S}{(|\psi \rangle _m}) - 6{\cal D}{(|\psi \rangle _m}{)^2} = 1,\\
{\cal S}{(|\psi \rangle _\theta }) + 3{\cal D}{(|\psi \rangle _\theta }{)^2} = 4 ,
\label{Eq.51}
\end{align}
which imply that states $|\psi \rangle _{m}$ are the upper boundary states for $0 \le {\cal D}(|\psi \rangle ) < 1/\sqrt3$ and the states $|\psi \rangle _{\theta}$ are the upper boundary states for $1/\sqrt3 \le {\cal D}(|\psi \rangle ) \le 1$.
\end{proof}

\begin{figure}
\centering
\includegraphics[width=8.6cm]{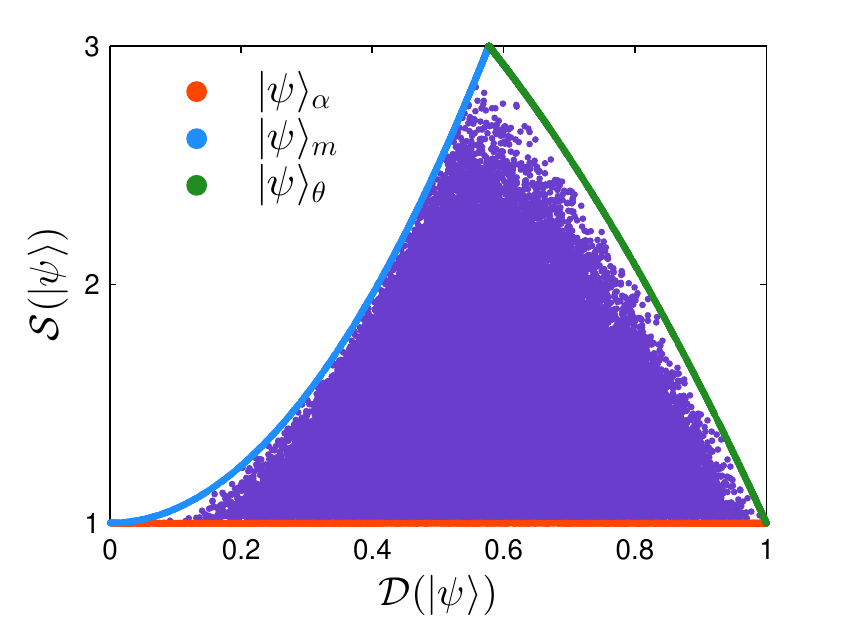}
\caption{(Color online) The close relation between the maximum steering inequality violation, ${\cal S}(|\psi \rangle )$, and the first-order coherence, ${\cal D}(|\psi \rangle )$, for ${10^5}$ Haar randomly generated three-qubit pure states. The state $|\psi \rangle _m$ lies at the left upper boundary, the state $|\psi \rangle _\theta$ lies at the right upper boundary, and the state $|\psi \rangle _\alpha$ lies at the $x$-axis. The $x$ and $y$ axes are dimensionless.}
\label{f5}
\end{figure}

In Fig. \ref{f5}, the relation between the maximum steering inequality violation and the first-order coherence is plotted for ${10^5}$ Haar randomly generated three-qubit pure states. The maximum steering inequality violation increases at first and then decreases with the increase of first-order coherence. The maximum steering inequality violation approaches its maximum 3 when the first-order coherence gets close to the critical value $1/\sqrt 3$. The states $|\psi \rangle _m$ and $|\psi \rangle _\theta$ lies at the left and right upper boundaries, respectively. The ${10^5}$ Haar randomly generated three-qubit pure states are
contained in the trilateral region.

\section{\label{sec6}SUMMARY}

In summary, we propose a general framework for unifying different measures of quantum resources, including genuine tripartite entanglement, coherence and quantum steering, for tripartite entanglement states.
First of all, the tradeoff relations between first-order coherence and genuine tripartite entanglement are established, where the genuine tripartite entanglement are quantified by GGM, GMC, and concurrence fill. There exists an exact functional form between GGM and GMC. The results show that the first-order coherence will be constrained to a range formed by two inequalities for a fixed amount of genuine tripartite entanglement. The upper boundary states of the tradeoff relation is state $|\psi \rangle _\alpha$, which possesses maximum first-order coherence value, and $|\psi \rangle _m$ is the lower boundary state. Moreover, we investigate the tradeoff relation between the maximum steering inequality violation and concurrence fill. Differently, in this case, the $|\psi \rangle _m$ state takes the maximum steering inequality violation for a given concurrence fill. In addition, we present the close
relation between the maximum steering inequality violation and first-order coherence. It is found that the upper boundary of the maximum steering inequality violation increase at first and then decrease with the increase of first-order coherence. When a critical value $1/\sqrt 3 $ of the first-order coherence is reached, the corresponding maximum steering inequality violation is its maximum 3.
The left upper boundary state is the state $|\psi \rangle _m$, and the right upper state is another boundary state $|\psi \rangle _\theta$. Both of them have the maximum steering inequality violation for a fixed value of first-order coherence. These results reveal that there exists a close connection among different measures of quantum resources in tripartite systems, which is of great significance to study the information transfer and flow in QRTs.

\begin{acknowledgements}
This work was supported by the National Science Foundation of China (Grant Nos. 12004006, 12075001, and 12175001), and Anhui Provincial Natural Science Foundation (Grant No. 2008085QA43).
\end{acknowledgements}

\appendix

\section{Supplementary proof of GMC versus GGM}
\hypertarget{A}{}
Here we give the proof of Eq. (\ref{Eq.cond1}). From the Eqs. (\ref{Eq.con1}) and (\ref{Eq.purity}), one can obtain
\begin{align}
{\rm{Tr}}\left( {\rho _A^2} \right) = \lambda _1^2 + \lambda _2^2 = 2\lambda _2^2 - 2{\lambda _2} + 1.
\label{Eq.A1}
\end{align}
Similarly, we get
\begin{align}
{\rm{Tr}}\left( {\rho _B^2} \right) = 2\lambda _4^2 - 2{\lambda _4} + 1,\quad{\rm{Tr}}\left( {\rho _C^2} \right) = 2\lambda _6^2 - 2{\lambda _6} + 1.
\label{Eq.A2}
\end{align}
Note that the right-hand sides of these equations take the same form of the function with $2{\lambda ^2} - 2\lambda  + 1$, which decreases monotonically in the interval that $\lambda\in[0,0.5]$. Since  ${\lambda _2}$, ${\lambda _4}$, and ${\lambda _6}$ are the smaller eigenvalues of the reduced density
matrices ${\rho _A}$, ${\rho _B}$, and ${\rho _C}$, respectively, we have $0 \le {\lambda _2},{\lambda _4},{\lambda _6} \le 0.5$. Assuming that ${\lambda _2} \le {\lambda _4}$ and ${\lambda _2} \le {\lambda _6}$, we get
\begin{align}
{\rm{Tr}}\left( {\rho _A^2} \right)\ge{\rm{  Tr}}\left( {\rho _B^2} \right),\quad {\rm{Tr}}\left( {\rho _A^2} \right)\ge{\rm{  Tr}}\left( {\rho _C^2} \right).
\end{align}
In addition, one can obtain
\begin{align}
 \frac{1}{2} \le Tr(\rho _i^2) \le 1,
\label{Eq.A4}
\end{align}
where $i \in \{ A,B,C\} $.

\section{Supplementary proof of first-order coherence versus GMC}
\hypertarget{B 1}{}
\subsection{Proof of Eq. (\ref{Eq.39a})}
To begin with, considering the inequality
\begin{align}
 {\rm{Tr}}\left( {\rho _B^2} \right) + {\rm{Tr}}\left( {\rho _C^2} \right) \le 2{\rm{Tr}}\left( {\rho _A^2} \right),
\label{Eq.B1}
\end{align}
we have
\begin{align}
  - \frac{4}{3}{\rm{Tr}}\left( {\rho _A^2} \right) + \frac{2}{3}{\rm{Tr}}\left( {\rho _B^2} \right) + \frac{2}{3}{\rm{Tr}}\left( {\rho _C^2} \right) \le 0.
\label{Eq.B2}
\end{align}
Then, it can be obtained that
\begin{align}
  1 - 2{\rm{Tr}}\left( {\rho _A^2} \right) + \frac{2}{3}{\rm{Tr}}\left( {\rho _A^2} \right) + \frac{2}{3}{\rm{Tr}}\left( {\rho _B^2} \right) + \frac{2}{3}{\rm{Tr}}\left( {\rho _C^2} \right) \le 1.
\label{Eq.B3}
\end{align}
Finally, we get
\begin{align}
 2\left[ {1 \!- \!{\rm{Tr}}\left( {\rho _A^2} \right)} \right] \!+\! \frac{2}{3}\left[ {Tr(\rho _A^2) \!+\! Tr(\rho _B^2) \!+\! Tr(\rho _C^2)} \right] \!-\! 1 \le 1.
\label{Eq.B4}
\end{align}

\hypertarget{B 2}{}
\subsection{Proof of Eq. (\ref{Eq.42a})}
To begin with, from the inequality
\begin{align}
 {\rm{Tr}}\left( {\rho _B^2} \right) + {\rm{Tr}}\left( {\rho _C^2} \right) \ge 1,
\label{Eq.B5}
\end{align}
we have
\begin{align}
 {\rm{ - 2Tr}}\left( {\rho _A^2} \right){\rm{ + 2Tr}}\left( {\rho _A^2} \right){\rm{ + 2Tr}}\left( {\rho _B^2} \right) + 2{\rm{Tr}}\left( {\rho _C^2} \right) \ge 2.
\label{Eq.B6}
\end{align}
Then, we can see that
\begin{align}
{\rm{2 - 2Tr}}\left( {\rho _A^2} \right){\rm{ + 2Tr}}\left( {\rho _A^2} \right){\rm{ + 2Tr}}\left( {\rho _B^2} \right) + 2{\rm{Tr}}\left( {\rho _C^2} \right) - 3 \ge 1.
\label{Eq.B6}
\end{align}
Finally, it gives
\begin{align}
2\left[ {1\! -\! {\rm{Tr}}\left( {\rho _A^2} \right)} \right] \!+\! 2\left[ {Tr(\rho _A^2) \!+\! Tr(\rho _B^2) \!+\! Tr(\rho _C^2)} \right] \!- \!3 \ge 1.
\label{Eq.B7}
\end{align}

\section{Supplementary proof of first-order coherence versus concurrence fill}
\hypertarget{C 1}{}
\subsection{Proof of Eq. (\ref{Eq.53a})}
Based on the inequality
\begin{align}
(Q - a)(Q - b)(Q - c) \le {\left( {\frac{Q}{3}} \right)^3},
\label{Eq.C1}
\end{align}
we have
\begin{align}
\frac{{16}}{3}Q(Q - a)(Q - b)(Q - c) \le 16{\left( {\frac{Q}{3}} \right)^4}.
\label{Eq.C2}
\end{align}
Then, we get
\begin{align}
{\left[ {\frac{{16}}{3}Q(Q - a)(Q - b)(Q - c)} \right]^{1/4}} \le \frac{2}{3}Q.
\label{Eq.C3}
\end{align}
Finally, we obtain
\begin{align}
\begin{split}
 {\left[ {\frac{{16}}{3}Q\left( {Q - a} \right)\left( {Q - b} \right)\left( {Q - c} \right)} \right]^{1/4}} + 1 - \frac{2}{3}Q \le 1.
 \end{split}
\label{Eq.C4}
\end{align}

\hypertarget{C 2}{}
\subsection{Proof of Eq. (\ref{Eq.56a})}
Using the inequality
\begin{align}
(2 - Q){(Q - 1)^2} \le (Q - a)(Q - b)(Q - c),
\label{Eq.C5}
\end{align}
we have
\begin{align}
\begin{split}
&\frac{{16}}{3}Q(Q - a)(Q - b)(Q - c) \ge \frac{{16}}{3}{(Q - 1)^2}(2 - Q)Q\\
 &  =  - \frac{1}{3}{(2Q - 2)^2} \cdot 4(Q - 2)Q\\
 &   =  - \frac{1}{3}{(2 - 2Q)^2}(4{Q^2} - 8Q)\\
 &    =  - \frac{1}{3}{(3 - 2Q - 1)^2}(4{Q^2} - 12Q + 9 + 4Q - 9)\\
 &     =  - \frac{1}{3}{\left[ {3\left( {1 - \frac{2}{3}Q} \right) - 1} \right]^2}\left[ {{{\left( {3 - 2Q} \right)}^2} - 6 + 4Q - 3} \right]\\
 &     =\!-{\left[ {3\left( {1 \!-\! \frac{2}{3}Q} \right) \!-\! 1} \right]^2}\left[ {3{{\left( {1\! -\! \frac{2}{3}Q} \right)}^2} \!- \!2\left( {1 \!- \!\frac{2}{3}Q} \right)\! - \!1} \right],
\end{split}
\label{Eq.C6}
\end{align}
Finally, we obtain
\begin{align}
\begin{split}
 &\frac{{16}}{3}Q\left( {Q - a} \right)\left( {Q - b} \right)\left( {Q - c} \right)\\
 &\! + \!{\left[ {3\left( {1 \!-\! \frac{2}{3}Q} \right) \!-\! 1} \right]^2}\left[ {3{{\left( {1\! -\! \frac{2}{3}Q} \right)}^2} \!- \!2\left( {1 \!- \!\frac{2}{3}Q} \right)\! - \!1} \right] \!\ge\! 0.
 \end{split}
\label{Eq.C7}
\end{align}

\hypertarget{D}{}
\section{Supplementary proof of the maximum steering inequality violation versus concurrence fill}
\hypertarget{D 1}{}
\subsection{Proof of Eq. (\ref{Eq.S43})}
From Eq. (\ref{Eq.C}), we have
\begin{align}
a \equiv C_{A(BC)}^2 = 4\det {\rho _A} = 4{\lambda _1}{\lambda _2},
\label{Eq.D1a}
\end{align}
Similarly, we find
\begin{align}
b = 4{\lambda _3}{\lambda _4},\quad c = 4{\lambda _5}{\lambda _6},
\label{Eq.D2}
\end{align}
Using the Eq. (\ref{Eq.60a}), one can obtain
\begin{align}
\begin{split}
{{\cal S}_{AB}}& = 4Tr(\rho _C^2) - 2Tr(\rho _A^2) - 2Tr(\rho _B^2) + 1\\
& = 4(\lambda _5^2 + \lambda _6^2) - 2(\lambda _1^2 + \lambda _2^2) - 2(\lambda _3^2 + \lambda _4^2) + 1\\
 & = 4(1 - 2{\lambda _5}{\lambda _6}) - 2(1 - 2{\lambda _1}{\lambda _2}) - 2(1 - 2{\lambda _3}{\lambda _4}) + 1\\
  & = 4{\lambda _1}{\lambda _2} + 4{\lambda _3}{\lambda _4} - 8{\lambda _5}{\lambda _6} + 1\\
  & = a + b - 2c + 1
\end{split}
\label{Eq.D3}
\end{align}

\hypertarget{D 2}{}
\subsection{Proof of Eq. (\ref{Eq.68b})}
Given
\begin{align}
Q = \frac{1}{2}(a + b + c),
\label{Eq.D4}
\end{align}
where $0 \le a,b,c \le 1$, then one can obtain that $a + b \le 2$. Then, we have
\begin{align}
4c \le 4 - 2a - 2b + 4c,
\label{Eq.D4}
\end{align}
This gives
\begin{align}
2b + 2c - 2a+2a + 2c - 2b   \le 4 - 2\left( {a + b - 2c} \right),
\label{Eq.D4}
\end{align}
The above equation can be rewritten as
\begin{align}
4\left( {\frac{a}{2} \!+ \!\frac{b}{2}\! +\! \frac{c}{2} \!-\! a} \right) \!+\! 4\left( {\frac{a}{2} \!+\! \frac{b}{2}\! +\! \frac{c}{2} \!-\! b} \right) \!\le\! 2\left[ {2\! - \!(a\! +\! b \!-\! 2c)} \right],
\label{Eq.D4}
\end{align}
Therefore, we obtain
\begin{align}
4(Q - a) + 4(Q - b) \le 2\left[ {2 - (a + b - 2c)} \right],
\label{Eq.D4}
\end{align}
\hypertarget{D 3}{}
\subsection{Proof of Eq. (\ref{Eq.70b})}
By using the inequality
\begin{align}
\begin{split}
4(Q - a)\cdot 4(Q - b)\cdot4(Q - c)\cdot4Q \le &{ [2 - (a + b - 2c)]^2}\\
\times\left( {2 + a + b - 2c} \right)&[6 - (a + b - 2c)],
\end{split}
\label{Eq.D6}
\end{align}
we have
\begin{align}
\begin{split}
&{16^2}Q(Q - a)(Q - b)(Q - c) \\
&\le  - {\left( {a + b - 2c - 2} \right)^2}\left( {a + b - 2c + 2} \right)\left( {a + b - 2c - 6} \right)\\
&= -{\left( {a \!+\! b \!-\! 2c \!+\! 1\! - \!3} \right)^2}\!\!\left( {a \!+ \!b \!-\! 2c \!+\! 1\! +\! 1} \right)\!\left( {a \!+\! b \!-\! 2c \!+\!1 \!-\! 7} \right).
\end{split}
\label{Eq.D7}
\end{align}
Thus, we get
\begin{align}
\begin{split}
&48 \times \frac{{16}}{3}Q\left( {Q - a} \right)\left( {Q - b} \right)\left( {Q - c} \right)\\
& \!+\! \!{\left( {a \!+\! b \!-\! 2c \!+\! 1\! - \!3} \right)^2}\!\!\left( {a \!+ \!b \!-\! 2c \!+\! 1\! +\! 1} \right)\!\left( {a \!+\! b \!-\! 2c \!+\!1 \!-\! 7} \right) \!\!\le\!\! 0.
\end{split}
\label{Eq.D8}
\end{align}

\section{Supplementary proof of the maximum steering inequality violation versus first-order coherence}
\subsection{\label{app:E1}Proof of Eq. (\ref{Eq.72a})}

Based on the inequality
\begin{align}
Tr(\rho _A^2) + Tr(\rho _B^2) \ge 1,
\label{Eq.E1}
\end{align}
we have
\begin{align}
 - 6Tr(\rho _A^2) - 6Tr(\rho _B^2) + 6 \le 0.
\label{Eq.E2}
\end{align}
Then, we can see that
\begin{align}
\begin{split}
4Tr(\rho _C^2) &- 2Tr(\rho _A^2) - 2Tr(\rho _B^2)\\
 &- 4Tr(\rho _A^2) - 4Tr(\rho _B^2) - 4Tr(\rho _C^2) + 6 \le 0.
 \end{split}
\label{Eq.E3}
\end{align}
Finally, we obtain
\begin{align}
\begin{split}
4Tr(\rho _C^2) &- 2Tr(\rho _A^2) -  2Tr(\rho _B^2) + 1 \\
&- 6\left[ {\frac{2}{3}\left( {Tr(\rho _A^2) +  Tr(\rho _B^2) + Tr(\rho _C^2)} \right) - 1} \right] \le 1.
\end{split}
\label{Eq.E4}
\end{align}
\subsection{\label{app:E2}Proof of Eq. (\ref{Eq.75a})}
From the inequality
\begin{align}
Tr(\rho _C^2) \le 1,
\label{Eq.E5}
\end{align}
we have
\begin{align}
6Tr(\rho _C^2)\! -\! 2Tr(\rho _A^2) \!-\! 2Tr(\rho _B^2)\! +\! 2Tr(\rho _A^2)\! +\! 2Tr(\rho _B^2) \!\le\! 6.
\label{Eq.E6}
\end{align}
Then, we can see that
\begin{align}
\begin{split}
4Tr(\rho _C^2) &- 2Tr(\rho _A^2) - 2Tr(\rho _B^2) + 1 \\
&+ 2Tr(\rho _A^2) + 2Tr(\rho _B^2) + 2Tr(\rho _C^2) - 3 \le 4.
\end{split}
\label{Eq.E7}
\end{align}
Finally, we obtain
\begin{align}
\begin{split}
4Tr(\rho _C^2) &- 2Tr(\rho _A^2) - 2Tr(\rho _B^2) + 1 \\
&+ 3\left[ {\frac{2}{3}\left( {Tr(\rho _A^2) + Tr(\rho _B^2) + Tr(\rho _C^2)} \right) - 1} \right] \le 4.
\end{split}
\label{Eq.E8}
\end{align}

%\bibliography{ref}


\end{document}